# Title: Transposable element sequence evolution is influenced by gene context

**Running head: Transposable element sequence evolution**

Anna-Sophie Fiston-Lavier[1,3], Charles E. Vejnar[2,3] and Hadi Quesneville[3]

[1] Department of Biology, Stanford University, Stanford, CA 94306, US.

[2] Computational Evolutionary Genomics group, Department of Genetic Medicine and Development, University of Geneva, 1211 Geneva 4, Switzerland.

[3] INRA, UR1164 URGI - Research Unit in Genomics-Info, INRA de Versailles-Grignon, Route de Saint-Cyr, Versailles, 78026, France.

Corresponding author:

Hadi Quesneville

INRA, UR1164 URGI - Research Unit in Genomics-Info, INRA de Versailles-Grignon,

Route de Saint Cyr, Versailles, 78026, France

Phone: +33 (0)1 30 83 30 08

Fax: +33 (0)1 30 83 34 58

hadi.quesneville [at] versailles.inra.fr

## Abbreviations:

TE        Transposable Element

LTR      Long Terminal Repeat

# Abstract

**Background:** Transposable elements (TEs) in eukaryote genomes are quantitatively the main components affecting genome size, structure and expression. The dynamics of their insertion and deletion depend on diverse factors varying in strength and nature along the genome. We address here how TE sequence evolution is affected by neighboring genes and the chromatin status (euchromatin or heterochromatin) at their insertion site.

**Results:** We estimated ages of TE sequences in *Arabidopsis thaliana*, and found that they depend on the distance to the nearest genes: TEs located close to genes are older than those that are more distant. Consequently, TE sequences in heterochromatic regions, which are gene-poor regions, are surprisingly younger and longer than that elsewhere.

**Conclusions:** We provide evidence for biased TE age distribution close or near to genes. Interestingly, TE sequences in euchromatin and those in heterochromatin evolve at different rates, and as a result, could explain that TE sequences in heterochromatin tend to be younger and longer. Then, we revisit models of TE sequence dynamics and point out differences for TE-rich genomes, such as maize and wheat, compared to TE-poor genomes such as fly and *A. thaliana*.

BACKGROUND

Transposable elements (TEs) are mobile, repetitive DNA sequences that constitute the most structurally dynamic components of the genomes of both plants and animals. TEs are virtually ubiquitous, and have been found in nearly all organisms studied. In many cases, TEs are the largest component of eukaryotic nuclear genomes: they constitute more than 50% of the human genome, 85% of the maize genome, and 88% of the wheat genome [1-4]. TEs have been shown repeatedly to be involved in genome restructuring, including the generation of inversions, translocations, and deletions [5-7]. It is now clearly established that TEs play a key role in genome structure and evolution.

Genome-wide TE analysis in plants and animals showed that TE amplification and elimination is very dynamic. TE sequences are efficiently eliminated from the genome, through both the accumulation of deletions and mutations, and ectopic recombination. Both the timing and extent of TE amplification differs between TE families. The turnover of TEs in genomes can be very high: for example, most of the TE sequences in rice were inserted recently, within the last two million years [8]. Some TE sequences such as LTR (Long Terminal Repeat) retroelements have undergone unequal homologous recombination events. This mechanism leads to the formation of TE sequences called *"solo-LTRs"* that eliminate large TE sequences. The half-life of such TE sequences is estimated to be less than three million years in rice [8].

TE insertions can be eliminated by purifying selection, when the genome rearrangements they induce (e.g., duplication, deletions) are deleterious. Moreover, TE insertions overlapping genes or regulatory regions, or close to genes, are also often deleterious, particularly if they disrupt regulation or function. In other cases, changes in expression due to TE insertion can be the source of adaptation [9-13]. TE insertions can epigenetically repress genes in their neighborhood [14]. In addition, we recently reported that DNA methylation can spread from a TE insertion to its neighborhood [15] possibly affecting expression of nearby genes. Indeed, Hollister and Gaut [9] showed that methylated TE sequences are associated with reduced expression of neighboring genes. This suggests that methylated TE sequences have stronger deleterious effects than unmethylated TE sequences on nearby genes. Consistent with this idea, the distances to neighboring genes are larger for methylated TE sequences than unmethylated TE sequences [9]. The proportion of unmethylated TE sequences close to the 5'-end of genes is higher than to the 3'-end suggesting that TE methylation may affect the promoters of nearby genes [15].

In *Arabidopsis thaliana,* ~20% of the genomic sequence is composed of TEs [15, 16]. If most TE insertions are considered deleterious, their insertion would be expected to be frequently followed by TE loss through natural selection [17]. Indeed, Wright *et al.* [10] showed that TE occurrence correlates negatively with gene density, in accordance with TEs having such a disruptive effect on nearby genes.

The negative impact of TE insertions on neighboring genes may explain in part the TE density, but may have other consequences. We therefore sought to determine whether the gene neighborhood affects their molecular sequence evolution. We also analyzed the relationship between the biology of the TE family (and/or superfamily) and TE sequence evolution. Exploiting the high quality repeat annotations of *A. thaliana*, our study provides the first assessment of the rates of TE sequence evolution based on a genome-wide analysis. We show that a key factor affecting TE sequence evolution in *A. thaliana* is their presence in gene-dense or gene-poor region also seen as euchromatin and heterochromatin respectively. We show that TE sequences in euchromatin and those in heterochromatin have evolved at different rates, and as a result, could explain that TE sequences in heterochromatin tend to be younger and longer.

## METHODS

### Data set

The *A. thaliana* genomic sequence was downloaded from the TAIR web site (http://www.arabidopsis.org). Repeat annotations were obtained with the "TEannot" pipeline in the "REPET "package [18, 19]. We previously annotated 31,245 TE sequences corresponding to 318 TE families [15, 16]. This annotation is available at the TAIR web site (http://www.arabidopsis.org). TEs constitute 21% (25 Mb) of the genomic sequence, and 3,342 sequences (0.85 Mb) are satellites. The segmental duplications were identified using the pipeline described by Fiston-Lavier *et al.* [7]. This pipeline allowed the identification of 1,930 repeats, clustered into 804 groups. Segmental duplications can be composed of other repeats. Thus to remove the redundancy, segmental duplications overlapping another repeat (*i.e.,* TE sequence or satellite) were removed such that only the no-TE and no-satellite regions were counted.

### Definition of euchromatin and heterochromatin domains

The limits of the heterochromatic domains on the five chromosomes of *A. thaliana* have been defined genetically, based on regions of reduced recombination [20, 21]. However they do not match well with the heterochromatin limits defined from epigenetic marks for chromosome 4 [22]. From this study, repeat density appears to be a good marker for the heterochromatin in *A. thaliana*. Hence, to refine these limits across the five chromosomes, we used TE and satellite densities. We calculated the cumulated coverage percentage along the pericentromeric and the knob regions. The cumulated sequence coverage was calculated for each window of 150 kilobases with an overlap of five kilobases between contiguous windows. This first step was used to define the threshold that optimally minimizes the difference between the genetic centromeric limits and our repeat-based estimated limits. The cumulated repeat coverage of some blocks in heterochromatic regions are slightly lower than the threshold identified, so we also defined a minimum spacer value to join the heterochromatic blocks.

To define heterochromatic domains, we used a minimum cumulated coverage value of 60% (TE and satellite density) and allowed the connection of heterochromatic blocks separated by less than one megabase. The comparison of repeat-based and genetic estimates [20, 21, 23] revealed only small differences (Supplemental Table S1). Based on the repeat density, the clear identification of the knob *hk5L* located on the long arm of the fifth chromosome was not accurate because of the low physical distance between this region and the pericentromere [24]. Therefore, we decided to gather annotations from *hk5L* and the pericentromere on the fifth chromosome, and to only distinguish the copies located in three genomic regions called hereafter "chromatin domains": euchromatin, heterochromatin, and *hk4S*.

### TE sequence alignments

We computed pairwise global nucleotide alignments of each TE sequence with the reference sequence from the family of the TE. For that, we used NWalign from the REPET package, implementing the Needleman and Wunsch algorithm [25]. The TE consensus used here as a reference sequence is generally considered as a proxy of the ancestral TE sequence, so only small indels (less than 100 basepairs) that did not occur in this consensus sequence are shown in the alignments. Consequently, the remaining indels correspond to true deletion events in the copies. The pairwise alignments were then used to compute multiple alignments, one per TE family.

### TE nucleotide divergence estimates and deletion analysis

TE divergences were estimated using three different approaches: (i) from the sequence divergence to the consensus sequence, (ii) from the sequence divergence between the LTR for LTR-retrotransposons (LTR-RTNs), and (iii) from the terminal branch forks length of TE copy phylogenies.

*Divergence from the sequence identity to the consensus sequence:* TE age can be approximated using the percent identity of the matches between the TE reference sequences and the annotated copies. Indeed, the consensus built from a multiple alignment of genomics copies following a simple majority rule at each base pair position, provides an approximation of their ancestral sequence. The observed divergence of each copy to this consensus, approximate the divergence since this ancestral copy. The divergence is directly obtained from the percent identity (the complement to 1) given by the TEannot pipeline when it aligns the genome to the TE reference sequences. This divergence can be converted into an insertion date by the use of a substitution rate, with the equation: $T=D/t$, where T is the time elapsed since the ancestral sequence, D the estimated divergence and t the substitution rate per site per year.

*Divergence from the sequence identity between the LTR for LTR-retrotransposons:* We estimated the insertion date of the complete LTR-RTN copies using the method based on the divergence between the two LTRs of the copy. Due to the LTR retrotransposon replication cycle, when a new copy inserts in the genome, its two LTRs are identical in sequence. As time elapses, the two LTR sequences accumulate mutations and thus diverge from each other. This divergence can be converted into an insertion date by the use of a substitution rate, with the equation: T=D/2t, where T is the time elapsed since the insertion, D the estimated LTR divergence and t the substitution rate per site per year. For each LTR-RTN copy we searched for LTR by aligning the two extremities using TRsearch (parameters -d 8 -g 16 -e 4 -i 0.6 -l 100), a tool from the REPET package [18, 19]. This divergence was calculated only for copies where at least 100bp could be aligned between the two LTRs over less that 500bp.

*Divergence from the terminal branch forks lengths:* We inferred trees, using PhyML [26], based on the multiple alignments obtained as described in the previous section. We selected pairs of aligned sequences from the same terminal branch fork to estimate their nucleotide divergence (*i.e.,* number of substitution per basepair). By doing this, we counted nucleotide substitutions between the closest TE sequence relatives pairs. These pairs could be then considered as mutually deriving from their last detectable transposition event. Hence, the number of differences between these sequences corresponds to only those substitutions that occurred after the sequence duplication leading to these two sequences (the TE transposition). This is a proxy of the TE copy age (expressed in substitution number per basepair) as they date the last transposition event. The terminal branch fork used is the one with two copies present in the reference genome. As recent transpositions may produce copies missing in the reference genome, some branch lengths may be longer than in reality and thus TE age overestimated. However, this bias is limited as TEs generally transpose by bursts. All copies transposing in the same time period, we may expect to have several copies of the same burst.

Considering TE sequences in terminal branch forks, led us to address 3,328 out of the 31,245 individual annotated TE insertions. Divergence is directly given by the branch length

The observed divergences were corrected for homoplasy using the Jukes and Cantor model and the substitution rate that we used was $1.3 \cdot 10^{-8}$ substitutions per site per year, as proposed by Ma and Bennetzen [1] after calibrating the substitution rate of the *adh* genes [2] to rice LTR retrotransposons.

*Deletion density estimate*: Deletion density (*i.e.,* number of deletion per basepair) was obtained directly from the alignment of sequence pairs in terminal forks by counting, with a customized script, the number of deletions and dividing it by the length of the considered TE sequence.

*Deletion rates estimates*: Deletion rate estimates per family or superfamily was obtained by the maximum likelihood approach proposed by Petrov *et al.* [27]. This method is based on the ratio of deletions to nucleotide substitutions. This approach is based on the assumptions that (1) the rates of deletion and substitution do not vary over time and (2) at any given time, the number of substitutions and deletions follow a "Poisson" distribution. A maximum likelihood estimator was calculated, as were confidence interval estimates with a $\chi^2$ approximation of the log-likelihood ratio.

Statistical analysis

Statistical analyses were performed with the R statistical package **(http://cran.r-project.org)**.

# RESULTS

TE sequence age correlates with gene distance

Most of TE insertions are expected to be deleterious. After insertion, TE sequences accumulate both point mutations and deletions. We analyzed the relationship between the substitution and deletion rates, and the distance to the closest gene.

Figure 1 shows TE ages estimates using three different methods (see Methods) for LTR retrotransposable elements (LTR-RT). Note that even if the three methods do not estimate exactly the same event, they appear to give consistent results for recent insertions (less than 1 Myr). However, we see that LTR estimates rely on a small number of copies (577) compared to the other, and appear to be biased toward recent events. This was expected as LTR-RT copies must keep the two LTRs intact to be used, and TE copies accumulate deletions with time, some of them removing one LTR. Age estimates using copy divergence to the consensus rely heavily on the quality of the consensus and then the exhaustivity of the aligned copies used to build them. Consequently, this approach may be biaised, and we preferred the terminal fork branch length estimate that appears to be more reliable as it does not suffer this potential bias. In addition, it could be used on all type of TEs, increasing the number of TEs being analysed.

Consequently, TE divergence were estimated using phylogenies of TE sequences. We normalized the values by (i) taking the square root of the nucleotide divergence and the log10 of the "deletion density" (*i.e.,* number of deletion per basepair), (ii) subtracting the mean value for the TE family, and (iii) dividing it by its standard deviation. Hence TE families can be compared on the same evolutionary scale.

Nucleotide divergence and deletion density decrease with distance to the nearest gene (Fig. 2A and 2B). Spearman's rank correlation tests are highly significant with *P-value* below 0.004 and 0.002 respectively for nucleotide divergence and deletion density (Both $\rho$ = -0.06). Wright *et al.* [10] reported that recombination does not correlate with TE abundance in *A. thaliana*, but there is a consistent negative correlation between gene density and TE abundance. They suggest that selection against ectopic recombination does not influence TE distribution, but is consistent with selection against TE disruption of gene expression.

Consequently, a simple explanation for the observed patterns could be that insertions into gene-rich regions are more likely to disrupt gene function and are therefore selected against reducing the likelihood of finding young TE insertions in gene-rich areas. Alternatively, this could be result of an accelerated rate of TE evolution close to neighboring genes. This could be a consequence of both (i) local disruption of gene expression by TE sequences and (ii) degenerate TE sequences restoring local gene expression being retained.

Interestingly, these findings for TEs in heterochromatic and euchromatic regions are very similar (Fig. 2A and 2B, respectively blue and green boxes), although the rates of recombination differ. Indeed, our definition of heterochromatin gives boundaries that correspond approximately to regions of suppressed recombination in *A. thaliana*. Therefore, recombination appears to make no contribution to the observed differences in age distribution.

These analyses confirm that gene density is a major factor affecting TE evolution as proposed by Wright *et al.* [10]. Euchromatin is typically gene-dense and heterochromatin relatively gene-poor.

Different patterns of DNA loss by small internal deletions

We investigated TE sequence evolution genome-wide and compared the features of TE sequences in euchromatic (*i.e.,* gene-rich) and heterochromatic (*i.e.,* gene-poor) regions. We were able to determine the limits with a reasonable accuracy for all five chromosomes of *A. thaliana* (see Methods; Supplemental Table S1; Supplemental Fig. S1). In addition to the pericentromeric regions, we also clearly re-identified an interstitial heterochromatic region: the knob called *hk4S* [20, 23]. The recent formation of this region suggests that there may be particular dynamics of the sequences located there [22]. We decided to differentiate it from the other heterochromatic regions.

Our analysis of gene presence in the neighborhood of a TE insertion also indicated that TE length decreases with decreasing distance to the nearest gene after insertion (Fig. 2C; Spearman's rank correlation tests: $\rho = 0.05$, *P-value* < 0.02). Therefore, we tested whether there are fewer and shorter deletions in heterochromatin, known to be gene-poor.

We estimated both the deletion rate in the two chromatic domains, and its 95% confidence intervals (see Methods). We used theses values to compare deletion rate estimates: when the confidence intervals do not overlap, we consider the estimations to be significantly different. Taking all sequences together, there was a significantly higher deletion rate for TE sequences in euchromatin than in heterochromatin (Table 1; 0.075 *vs.* 0.063 deletion per basepair). This is in good agreement with our previous findings showing that deletion density is correlated with gene proximity.

TEs are subdivided into subfamilies, families, superfamiles, orders, subclasses and classes [28]. Here, we compared the estimates between superfamilies and/or groups of superfamilies categorized as such having similar TE family biology (*i.e.,* transposition mechanisms and ranges of sequence length). For some of these categories the confidence intervals estimated for euchromatin and heterochromatin overlap indicating no significant difference in the deletion rates. Note, however, that the statistical power of some of these analyses is low because of the small number of deletion events in the corresponding sub-categories (Table 1). Interestingly, the rates differ between the categories consistent with the existence of TE superfamily-specific deletion mechanisms. Indeed, Gypsy-like superfamily shows an increased rate of deletion in heterochromatin, indicating such a superfamily specific mechanism. The deletion rate measured here may reflect a combination of both genomic and TE-specific deletion mechanisms. Also, deletions in TE sequences appear to be generally longer in heterochromatin (Table 1), so it is plausible that different deletion mechanisms apply in different chromatin domains.

The effect of Intra-chromosomal recombination

Deletions may also be induced by intra-chromosomal homologous recombination between TE copies. This mechanism generally eliminates large genomic segments, and not only the TEs themselves. Thus, on top of mechanisms inducing short deletions, intra-chromosomal recombination may contribute to TE sequence removal, and do so rapidly.

Solo-LTR sequences are the result of such recombination between the two LTR regions of one retroelement. The solo-LTR fraction is thus proportional to this recombination rate. We therefore estimated the fraction of solo-LTRs in the two chromatic domains as measure of the relative strength of the intra-chromosomal recombination mechanism. Based on coordinates in the reference TE, obtained during the annotation, we identified the solo-LTR copies and compared their number to that of other TE copies from the same family (Table 2). Most LTR elements were truncated at the 5' and/or 3' termini. There was also enrichment of truncated Copia-type copies in euchromatin whereas full Gypsy-type copies were over-represented in the heterochromatin (163 sequences vs. 83 in euchromatin; Table 2). These results are in agreement with the observation of longer TE sequences in the heterochromatin. Nevertheless, solo-LTR fractions for the two superfamilies combined did not differ between the heterochromatin and the euchromatin. This suggests a similar intra-chromosomal recombination rate in the two chromatic domains. However, because of the higher TE density in heterochromatin, more recombination events would be expected between TE sequences in this domain.

More divergent TE sequences in euchromatin

We aligned each TE copy and its reference sequence, and calculated the percentage of identity. The resulting identity distribution per TE family illustrates divergence between copies related to their transposition history. These distributions indicate waves of TE invasion suggesting multiple TE invasions by "bursts" of transpositions (*i.e.,* numerous transposition events over a short period of time); there were more than one burst for each TE family (Supplemental Fig. S2).

The TE sequence identity varied between 58% and 100% with a mean at 80.4%. Mean identity is related to the age of the TEs: high identity scores indicate that the sequences have accumulated very few mutations since their divergence from there common ancestor (TE consensus used here as a reference can be considered as a proxy of the ancestral sequence state) and therefore can be considered to be young. The mean identity did not differ between the heterochromatic and the knob TE sequences (mean sequence identity: 81.9% in heterochromatin and 82% in the knob; Welch two sample t-test: $t = 0.14$, $df = 131.18$, *P-value* = 0.89). However, euchromatic TE sequences had a lower mean identity of 79.9% (Welch two sample t-test: $t = 21.91$, $df = 13307.17$, *P-value* < 0.001) and thus appear to be older than heterochromatic TE sequences. We observed more diverged TE sequences in gene-rich regions than in gene-poor regions.

An alternative and non-exclusive explanation to the increased deletion rate in euchromatin, is that intra-chromosomal recombination, preferentially involving young TE copies because of their higher sequence similarity, may quickly eliminate older TE sequences located between the young TE copies. Whether this is the case, we would expect to observe more young TEs in heterochromatin, because of its high TE density leading to a higher rate of recombination of this type and consequently a higher TE removal rate.

Interestingly, this trend was not observed for all the TE superfamilies, suggesting possible effects associated with TE superfamily biology. For example, TE copies from the *HAT*, *En-Spm, Harbinger, L1, SINE* and *Copia* superfamilies appear younger in euchromatin than heterochromatin (Fig. 3).

Evolution is affected differently according to TE superfamily.

Some TE families show insertion biases [29-33]. This could affect the global rate of evolution of the family if the insertion bias is related to gene location. Consequently, even if sequence evolution within each superfamily was similar in the chromatin domains, different mechanisms would be expected to drive TE sequence evolution for different superfamilies.

For all TE sequences, except for the *En-Spm* and *Gypsy-type* superfamilies, we observed a small bias of accumulation in euchromatin relative to heterochromatin (Fig. 4). The opposite bias was detected only for the *Gypsy-type* elements (70.7% of the copies in the heterochromatin; $\chi^2$ = 580.6, df = 1, *P-value* << 0.001). Combining age (Fig. 3) and distribution bias information, our results indicate a clear insertion preference for *Gyspy-type* TE copies in the heterochromatin, consistent with previous observations [30, 33]. As TEs from *Gypsy-type* and *En-Spm* superfamilies insert preferentially into gene poor regions, they would be expected to evolve more slowly than TEs from other superfamilies.

Few TE superfamilies are represented in the *hk4S*. Most of the TE sequences in this region are members of the highest copy-number families (e.g., *Helitron*, LINE, *Gypsy-type*, *Copia-type* and *MuDR*). The TE distribution profile in the *hk4S* was intermediate between those of the euchromatin and the heterochromatin, supporting the apparently special status of the knob (Fig. 4).

# DISCUSSION

## Model of TE population dynamics

We report the first genome-wide analysis estimating the rates of TE evolution. We exploited the high-quality TE annotations available in *A. thaliana* to compare our results to those expected from two current and non-exclusive models of TE population dynamics: the "ectopic recombination" and the "gene-disruption" models [34].

Because they are present as dispersed homologous sequences, TEs may induce ectopic recombination leading to genome rearrangements (e.g., duplications, deletions and inversions). The "ectopic recombination" model suggests that TE sequences are eliminated due to the deleterious effects of these genome rearrangements. By consequence, TE sequences in high-recombination rate regions will be more quickly eliminated [35, 36]. This model also predicts an accumulation of TE sequences in regions with low meiotic recombination rates [37]. Indeed, in *Drosophila melanogaster*, the accumulation of TE sequences is negatively correlated with recombination rates, consistent with recombination contributing to TE elimination [38-41]. However, in selfing species, ectopic recombination is believed to be rare, as selfing induces homozygosis. TE insertions are homozygous, so for a given allelic position, template choice during the recombination process will be driven towards the allelic position on the sister chromatid or the homologous chromosome, preventing ectopic homologous repair. The effect of recombination on TE distribution in selfing species is thus expected to be weak [37]. Consistent with this notion, no significant correlation between the TE density and the recombination rate in *A. thaliana* or in *Caenorhabditis elegans*, were found, where selfing is frequent [10, 42]. According to the "gene-disruption" model, TE insertions into genes or regulatory regions are strongly selected against. Consequently, repeats accumulate in gene-poor regions [12, 43]. Similarly to what was reported by Wright *et al.* [10], we observed a significant inverse correlation between the densities of repeats and gene in *A. thaliana* (data not shown), indicating that the presence of TEs within or close to genes is deleterious. In addition, our observations on TE sequence evolution strongly support the "gene-disruption" model, as it explains the higher rate of TE evolution close to genes. This model appears to provide a good explanation of TE dynamics in the *A. thaliana* genome and perhaps also in most selfing species.

Our results also provide clear evidence of insertion bias of *Gypsy-type* copies in favor of the heterochromatin [30]. These biases have already been observed for LTR retroelements in other species, notably wheat and yeast [29, 44]. This insertion bias seems to be induced by specific interactions between LTR retroelements and proteins associated with heterochromatin [33].

TE sequence evolution

Our results suggest that gene density drives not only the dynamics of TE insertion but also TE sequence evolution. TE sequences close to genes are more divergent and shorter. TE sequence methylation may be involved because it could affect nearby genes. Ahmed *et al.* [15] showed that methylation might spread around TE sequences, propagating the repression state in their flanking regions. Consequently, when a TE sequence is close to a gene, silencing the gene expression, it may be selected against. In *Arabidopsis,* siRNA are believed to target TE sequences, due to sequence similarities, and induce methylation [15, 45]. TE sequences could escape this repression by diverging from the siRNA pool produced by other

TE copies; the resulting degenerated TE sequences close to genes would then be better tolerated. This is entirely consistent with our observations.

Within regions of high gene density, selection pressure on gene expression would induce a faster elimination of TE insertions. TE removal may be driven by segregating haplotypes with no TE insertion or, if the TE insertion is frequent in the population, by preserving shorter TE sequences. In gene-poor regions, there may be almost no selection pressure and long TE sequences could persist. Here, TE sequences are eliminated through the accumulation of short deletions or by ectopic non-meiotic recombination events, in a quasi-neutral way. Consequently, TE sequences accumulate in heterochromatin because of its low gene density. Note, that heterochromatin was often considered to be a cemetery for TEs, where TE sequences are old and degenerate compared to those in euchromatin. Here, we observe the reverse: TE copies appear to be longer and younger in heterochromatin than in euchromatin.

We suspect that one mechanism of sequence elimination has been underestimated: large deletions caused by ectopic recombination between young TE sequences (that flip out large DNA fragments full of TE sequences). This mechanism is expected to be more important in heterochromatic regions, as the density of young repeats is higher and genes lower. Removal of large DNA fragments affecting a gene will be strongly selected against and may indeed not be observable at the population level. We found no evidence that ectopic recombination rates, estimated by examining *solo-LTR* formation as a proxy, differed between euchromatin and heterochromatin. Note that *solo-LTRs* are the result of unequal intra-element homologous recombination between two LTR. This event occurs over a short distance and does not involve any host genes that could lead to selection against the recombination. As a consequence, this proxy does not appear to provide a good estimate of the global ectopic recombination rate.

### TE-rich and TE-poor genome dynamics

Our findings provide insight into various observations made on TE-rich genomes, particularly maize and wheat with 85% and 88% [1-3], respectively, of their genomes made of TEs. They also allow some interesting predictions about their sequence evolution. In these genomes, gene density is expected to be relatively low. Consequently, TE sequences may evolve almost without any gene selective pressure. We therefore expect to observe longer TE sequences with only limited divergence between copies of a TE family in TE-rich genomes. The proportion of full-length copies is expected to be higher in such genomes than in TE-poor genomes. Because of the longer TE sequences and the higher number of repeats, the genome is expected to be larger with large regions full of repeats, genes being dispersed in these regions with a tendency to form gene islands to escape epigenetic TE repression of the surrounding sequences. In these genomes, TE copies are mainly eliminated by ectopic recombination, a mechanism less efficient in selfing species. Their genome would increase in size faster than other genome. In TE-poor genomes, TE copies are expected to be short and heavily fragmented, and the divergence between copies of a TE family to be higher than in a TE-rich genome. The gene selective pressure would shape the neighborhood of the genes to escape the epigenetic repression of nearby TE copies, as explained above.

The transition from a TE-poor genome to a TE-rich genome appears to be a one-way process. Indeed, if one or several TE families successfully invade a TE-poor genome and reduce the gene density, TE copies would accumulate more and more efficiently. The gene selective pressure would decline as the gene density decreases and TE elimination will therefore become less efficient. TE copies would first accumulate in regions devoid of genes, which would then grow. These regions would be generally located near the centromere. The pericentromeric regions would expand, pushing genes towards the chromosome extremities, as observed in wheat [2]. Epigenetic control of the TE sequences would induce, in these regions, the formation of heterochromatin that in turn would suppress meiotic recombination. The Hill-Robertson [46] effect being less efficient, genes will be selected against and would therefore tend to disappear from TE-rich domains. In this state, it would be difficult for the genome length to decrease. TE-rich regions being no longer subject to selective pressure would evolve neutrally. When a large deletion occurs, genetic drift alone would determine its success in invading the population. According to standard population genetic theory, the deletion will be most often lost from the population as its frequency would be $1/(2N_e)$ (with $N_e$ the effective population size).

The *hk4S* knob is the result of a recent "heterochromatinisation" event where there is an accumulation of TE sequences relative to the homologous region on the long fourth chromosome arm [22]. It has conserved eight of the 33 genes of the donor region in the same order and orientation. Because the gene density is higher in the *hk4S* knob than in heterochromatin, we expect the gene selection pressure to be higher than that in heterochromatin but lower than that in euchromatin. The patterns of TE copies within this region show indeed an intermediate profile between those of euchromatin and heterochromatin. This may be an illustration of the transition from TE-poor towards TE-rich status.

## Conclusions

Exploiting the high quality TE annotations in *A. thaliana*, we present here the first genome-wide assessment of TE evolution. We provide evidence for accelerated rates of sequence evolution in gene-rich regions. Consequently, TE sequences in euchromatin and those in heterochromatin evolve at different rates, and as a result, TE sequences in heterochromatin tend to be younger and longer. This result suggests that TE evolve at a different rate according to genome gene density.

## Authors' contributions

HQ conceived the study. AF, CV, and HQ performed analysis. AF and HQ wrote the manuscript.

## Acknowledgments

We thank Ruth Hershberg, Florian Maumus, Dmitri Petrov and Diamantis Sellis for helpful discussions and comments on earlier drafts of this manuscript. This work was supported by the Institut Nationale de la Recherche Agronomique (INRA) and the Centre National de la Recherche Scientifique (CNRS) Groupement de Recherche "Elements Transposables" [to H.Q.]

**Figure 1**

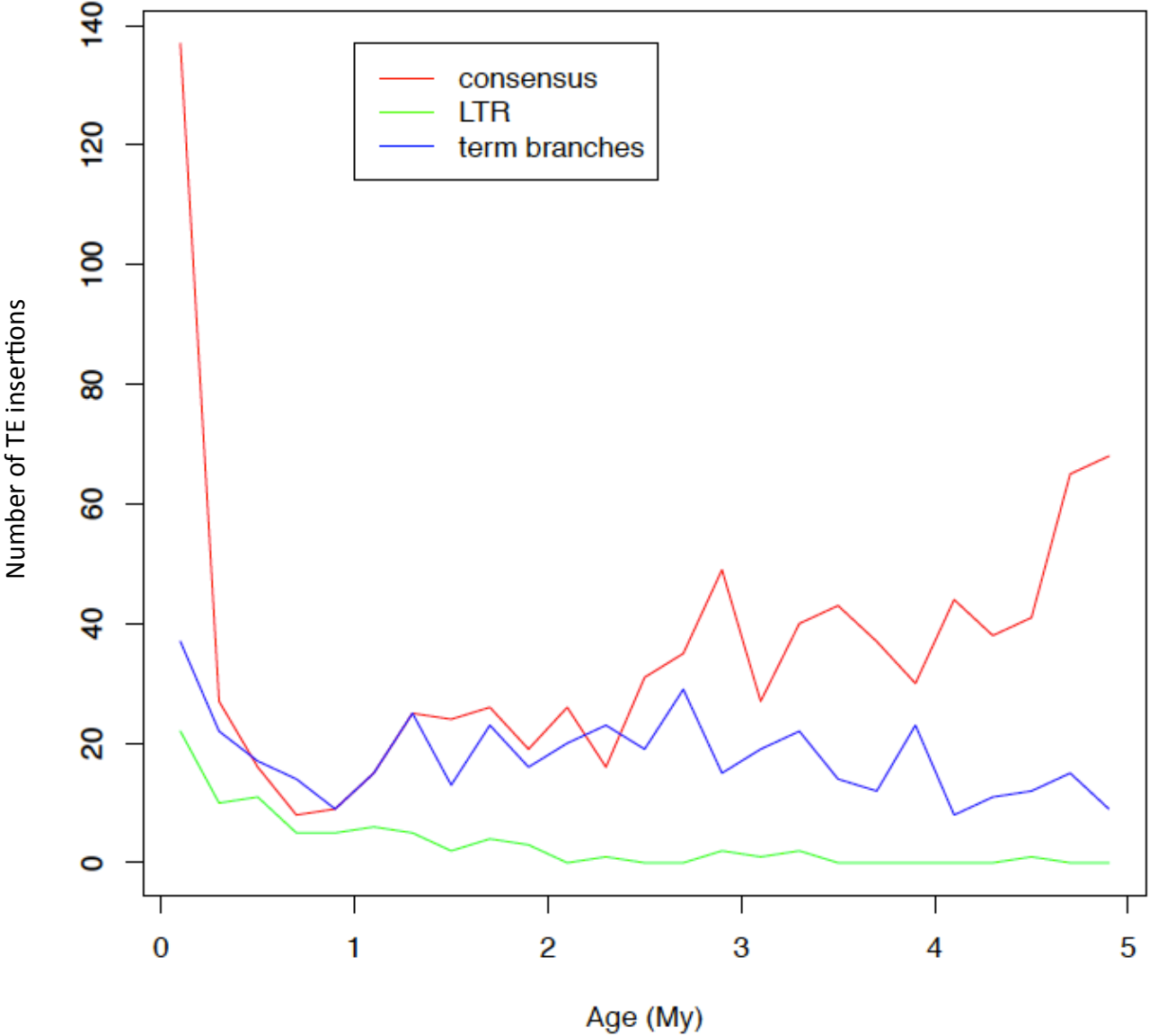

# Figure 2

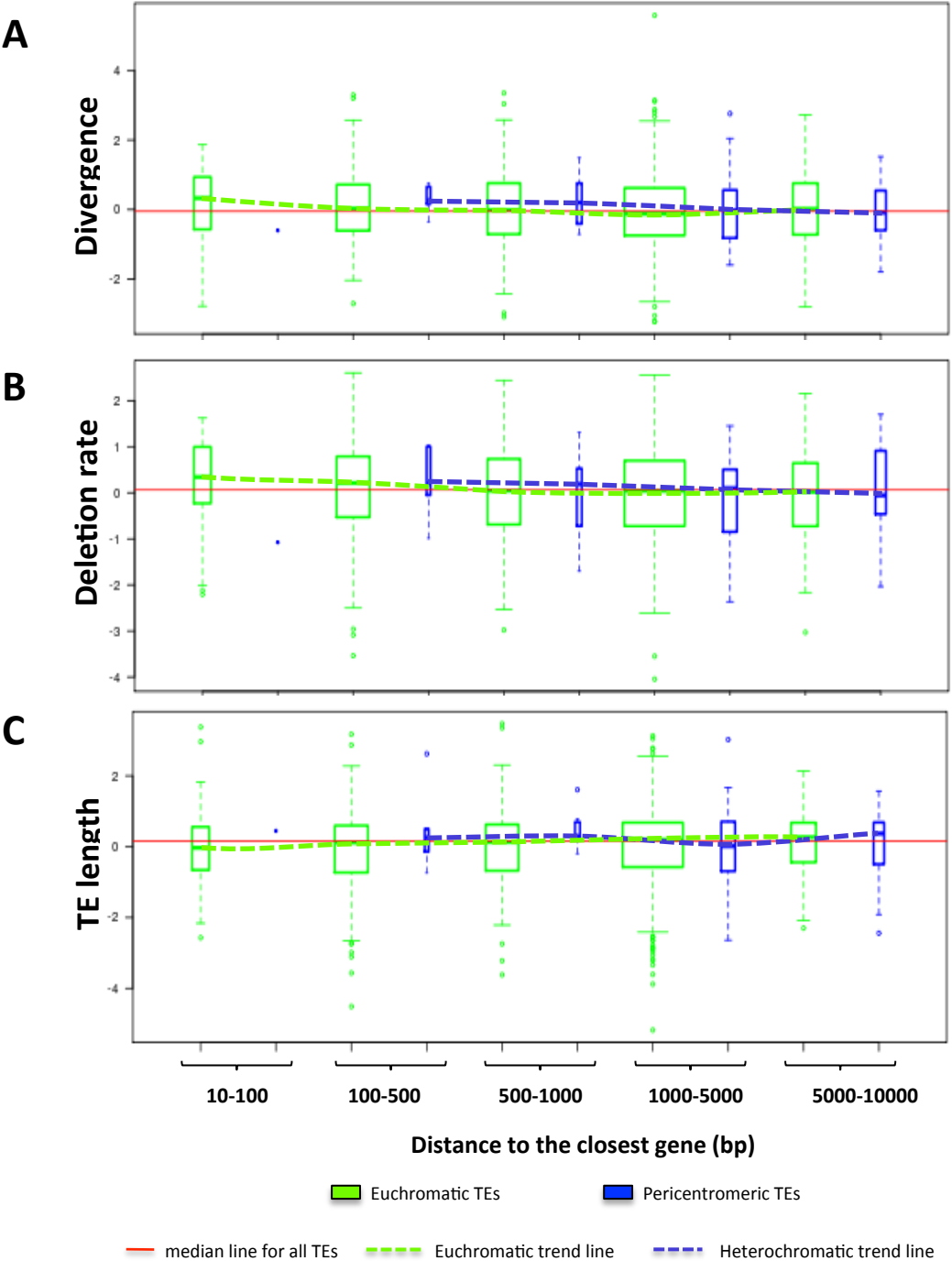

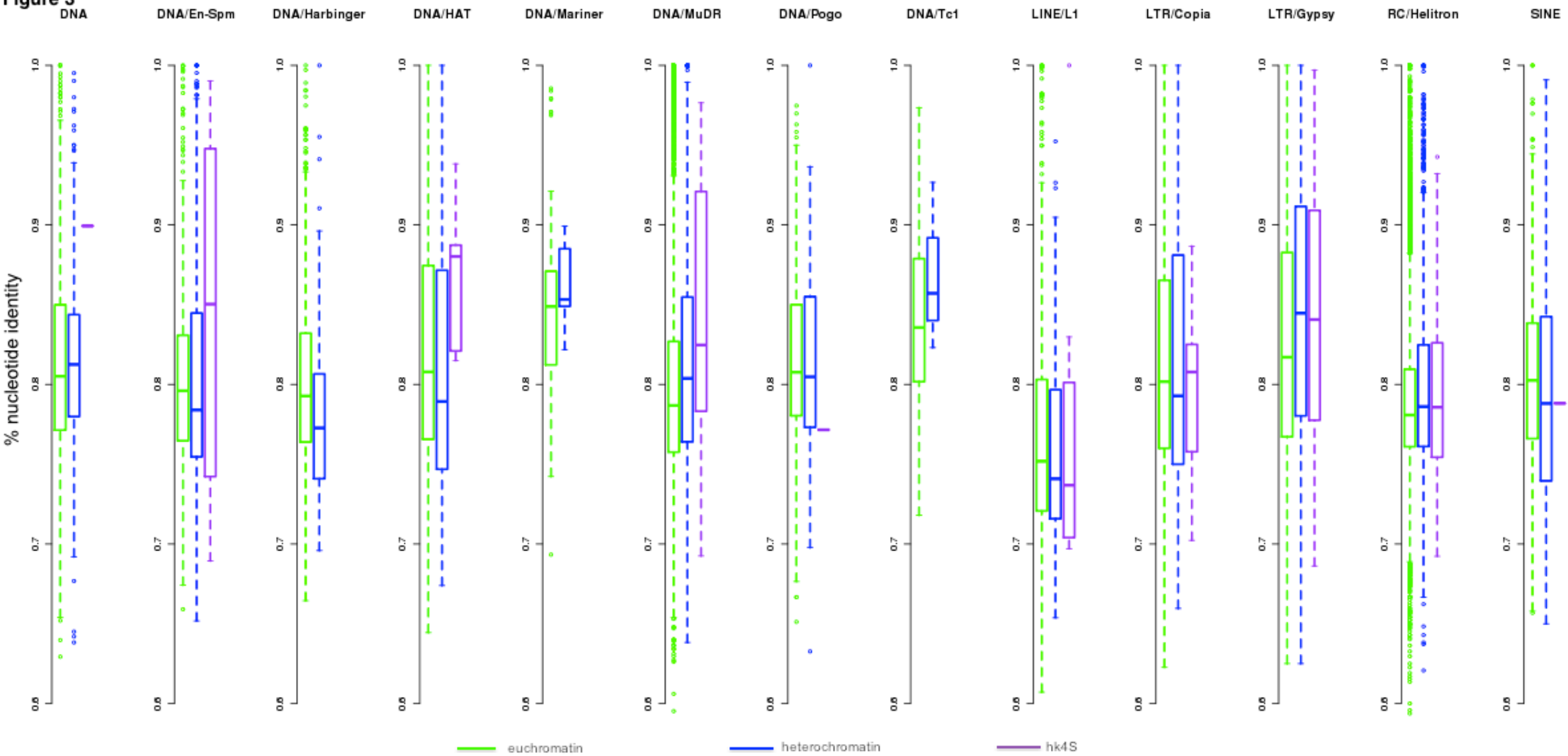

Figure 3

Figure 4

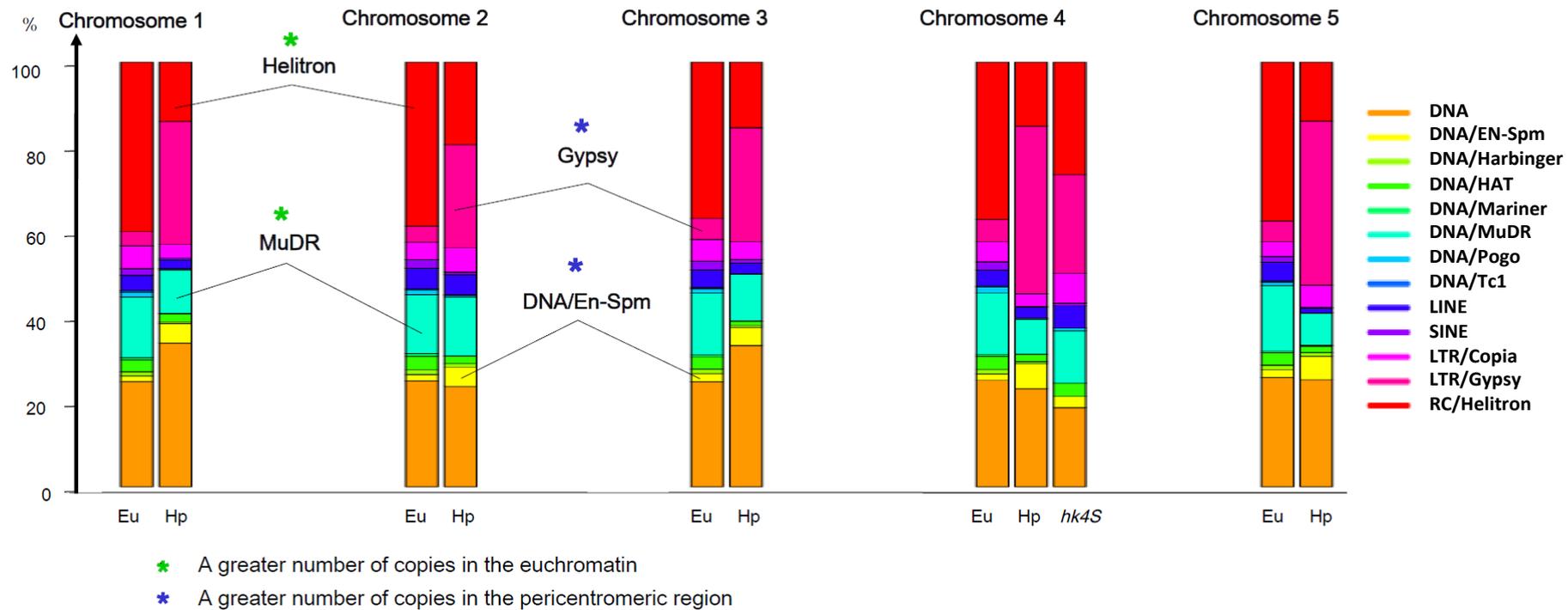

## Figure and table legends

**Figure 1.** Age estimates of LTR retrotransposons using three different methods. Red: age estimated from the divergence to the consensus; Blue: age estimated from the terminal forks branch length; Green: age estimated from the divergence between the two LTR of a copy.

**Figure 2.** Effect of the nearest gene on TE sequence evolution. Each value is normalized (i) taking the square root of the divergence and the log10 of the deletion rate or the TE length, (ii) subtracting the mean value for the TE family, and (iii) dividing it by its standard deviation. Values are grouped by distance classes to the nearest gene. The green and the blue boxplots correspond to the euchromatic and heterochromatic TEs, respectively. Box widths are proportional to the number of values in each class. The two edges of a box correspond to the first and the third quartiles. The horizontal bold line of the box represents the median (or second quartile). The "mustaches" of each boxplot are placed respectively above and below the third and first quartile at 1.5 times the first-third interquartile range. Values outside the "mustaches" range are considered as outliers, and are represented as dots. The red line indicates the median for all the TEs without chromatin domain distinction. A) Divergence distribution. B) Distribution of the deletion rate. C) Distribution of the length of the TE sequences.

**Figure 3.** Distribution of the TE divergence per superfamily and per chromatin domain. For each superfamily, we plotted the three TE divergence distributions using boxplot

representations, one for each chromatin domain. The green, blue and purple boxplots correspond to euchromatin, heterochromatin and the knob *hk4S*, respectively. The two edges of a box correspond to the first and the third quartiles, and the horizontal line inside the box represents the median (or second quartile). The "mustaches" of each boxplot are placed respectively above and below the third and first quartile at 1.5 times the first-third interquartile range. Values outside the "mustaches" range are considered as outliers, and are represented as dots.

**Figure 4.** TE proportion according to superfamily and chromatin domain.

**Supplemental Figure S1.** Cumulated sequence coverage along the five chromosomes.

The coverage of TEs, gene duplications and satellites are represented in green, purple and red, respectively. The gene (non-duplicated) coverage is in blue. The yellow lines indicate the centromere location. The black and red lines correspond, respectively, to the heterochromatin genetic and estimated limits. The genetic data are from Fransz *et al.* [20]. TEs and satellites are overabundant in heterochromatin and in the hk4S. In contrast, gene duplications are interspersed along the chromosomes.

**Supplemental Figure S2.** TE divergence distribution according to family.

**Table 1.** Deletion rate estimates according to superfamily group and chromatin domain.

| Superfamily/ Groups of superfamilies | Euchromatic regions | | | Pericentromeric regions | | |
|---|---|---|---|---|---|---|
| | Deletion rate [IC95%] | Mean deletion size [IC95%] | Mean copy size | Deletion rate [IC95%] | Mean deletion size [IC95%] | Mean copy size |
| **All DNA TEs** | 0.094 [0.086-0.091] | 19 [15-23] | 3063 | 0.085 [0.079-0.088] | 19 [15-24] | 2714 |
| **All LINEs** [1] | 0.058 [0.054-0.061] | 11 [10-11] | 3067 | 0.050 [0.045-0.054] | 12 [10-13] | 3126 |
| **SINE** | 0.087 [0.072-0.094] | 6 [5-8] | 420 | 0.11 [0.055-0.185] | 3 [1-6] | 460 |
| **LTR/Copia-type** | 0.042 [0.039-0.045] | 17 [12-23] | 3451 | 0.054 [0.044-0.064] | 22 [9-40] | 3594 |
| **LTR/Gypsy-type** [1] | 0.050 [0.046-0.053] | 22 [16-28] | 4014 | 0.055 [0.052-0.056] | 37 [31-45] | 5130 |
| **RC/Helitron** [1] | 0.089 [0.083-0.090] | 14 [13-15] | 2480 | 0.071 [0.062-0.077] | 20 [16-26] | 2902 |
| **All TEs** [1] | 0.075 [0.071-0.073] | 16 [15-18] | 3002 | 0.063 [0.060-0.063] | 26 [22-30] | 3076 |

(1) Significant difference between the two chromatic domains.

**Table2.** Distribution of different types of LTR retroelements by superfamily and by chromatic domain.

| % domain (number) | full-LTR[1] | | LTR-IR[2] | | IR[3] | | solo-LTR[4] | |
|---|---|---|---|---|---|---|---|---|
| | Eu[5] | PH[6] | Eu[5] | PH[6] | Eu[5] | PH[6] | Eu[5] | PH[6] |
| **Gypsy-type** | 10.49 (83) | 14.11 (163) | 35.52 (281) | 43.03 (497) | 43.24 (342) | 31.52 (364) | 13.02 (103) | 11.34 (131) |
| **Copia-type** | 15.37 (142) | 12.93 (30) | 34.42 (318) | 30.60 (71) | 43.24 (342) | 48.28 (112) | 9.85 (91) | 8.19 (19) |

(1) Full-length elements with the IR (Internal Region) and the two LTRs.
(2) Elements composed of the IR (Internal Region) and one of the two LTRs.
(3) Elements composed of only the IR (Internal Region).
(4) Elements resulting from the recombination between two LTRs, composed of one chimeric LTR.
(5) Euchromatin.
(6) Pericentromeric Heterochromatin.

**Supplemental Table S1.** Estimates of the heterochromatin limits for the five chromosomes. In the heterochromatin, average TE and satellite coverage is 70.7% and 5.7%, respectively; genes are more abundant in the euchromatic domains with an average coverage of 40.9%. Satellite coverage is higher in the *hk4S* than the heterochromatin (20% against 5.7%; χ2 = 7.9, df = 1, *P-value* < 0.01). This explains the higher satellite density in the fourth and fifth chromosomes containing a knob, (47.4 and 42 satellites/Mb vs. 36.1, 16.6 and 4.9 satellites/Mb for the first, the second and the third chromosomes, respectively).

| Heterochromatic domain | Chromosome | Genetic heterochromatic limits [1] | | Estimated heterochromatic limits | |
|---|---|---|---|---|---|
| | | start | end | start | end |
| Pericentromeric heterochromatin | Chromosome 1 | 15,088,987 | 15,605,649 | 13,415,000 | 16,025,000 |
| | Chromosome 2 | 3,608,427 | 3,852,785 | 2,395,000 | 6,020,000 |
| | Chromosome 3 | 13,810,904 | 14,507,140 | 12,110,000 | 15,155,000 |
| | Chromosome 4 | 3,956,519 | 4,206,698 | 3,120,000 | 5,150,000 |
| | Chromosome 5[2] | 1,742,755 | 12,621,612 | 10,660,000 | 13,270,000 |
| Knob (hk4S) | Chromosome 4 | 1,600,000 | 2,330,000 | 1,670,000 | 1,960,000 |

(1) Data from Fransz *et al.* [20].
(2) A knob domain has been also identified on this chromosome but the short distance from this domain to the pericentromeric domain makes it difficult to distinguish between them.

# Literature cited

1. Schnable PS, Ware D, Fulton RS, Stein JC, Wei F, Pasternak S, Liang C, Zhang J, Fulton L, Graves TA, et al: **The B73 maize genome: complexity, diversity, and dynamics.** *Science* 2009, **326:**1112-1115.
2. Choulet F, Wicker T, Rustenholz C, Paux E, Salse J, Leroy P, Schlub S, Le Paslier MC, Magdelenat G, Gonthier C, et al: **Megabase level sequencing reveals contrasted organization and evolution patterns of the wheat gene and transposable element spaces.** *Plant Cell* 2010, **22:**1686-1701.
3. Lander ES, Linton LM, Birren B, Nusbaum C, Zody MC, Baldwin J, Devon K, Dewar K, Doyle M, FitzHugh W, et al: **Initial sequencing and analysis of the human genome.** *Nature* 2001, **409:**860-921.
4. de Koning AP, Gu W, Castoe TA, Batzer MA, Pollock DD: **Repetitive elements may comprise over two-thirds of the human genome.** *PLoS Genet* 2011, **7:**e1002384.
5. Wicker T, Buchmann JP, Keller B: **Patching gaps in plant genomes results in gene movement and erosion of colinearity.** *Genome Res* 2010, **20:**1229-1237.
6. d'Alencon E, Sezutsu H, Legeai F, Permal E, Bernard-Samain S, Gimenez S, Gagneur C, Cousserans F, Shimomura M, Brun-Barale A, et al: **Extensive synteny conservation of holocentric chromosomes in Lepidoptera despite high rates of local genome rearrangements.** *Proc Natl Acad Sci U S A* 2010, **107:**7680-7685.
7. Fiston-Lavier AS, Anxolabehere D, Quesneville H: **A model of segmental duplication formation in Drosophila melanogaster.** *Genome Res* 2007, **17:**1458-1470.
8. Vitte C, Panaud O, Quesneville H: **LTR retrotransposons in rice (Oryza sativa, L.): recent burst amplifications followed by rapid DNA loss.** *BMC Genomics* 2007, **8:**218.
9. Hollister JD, Gaut BS: **Epigenetic silencing of transposable elements: a trade-off between reduced transposition and deleterious effects on neighboring gene expression.** *Genome Res* 2009, **19:**1419-1428.
10. Wright SI, Agrawal N, Bureau TE: **Effects of recombination rate and gene density on transposable element distributions in Arabidopsis thaliana.** *Genome Res* 2003, **13:**1897-1903.
11. McDonald JF, Matyunina LV, Wilson S, Jordan IK, Bowen NJ, Miller WJ: **LTR retrotransposons and the evolution of eukaryotic enhancers.** *Genetica* 1997, **100:**3-13.
12. Finnegan DJ: **Transposable elements.** *Curr Opin Genet Dev* 1992, **2:**861-867.
13. Gonzalez J, Lenkov K, Lipatov M, Macpherson JM, Petrov DA: **High rate of recent transposable element-induced adaptation in Drosophila melanogaster.** *PLoS Biol* 2008, **6:**e251.
14. Martienssen RA: **Heterochromatin, small RNA and post-fertilization dysgenesis in allopolyploid and interploid hybrids of Arabidopsis.** *New Phytol* 2010, **186:**46-53.
15. Ahmed I, Sarazin A, Bowler C, Colot V, Quesneville H: **Genome-wide evidence for local DNA methylation spreading from small RNA targeted sequences in Arabidopsis**

. In *Book Genome-wide evidence for local DNA methylation spreading from small RNA targeted sequences in Arabidopsis* (Editor ed.^eds.). City; 2011.
16. Buisine N, Quesneville H, Colot V: **Improved detection and annotation of transposable elements in sequenced genomes using multiple reference sequence sets.** *Genomics* 2008, **91:**467-475.
17. Lockton S, Gaut BS: **The evolution of transposable elements in natural populations of self-fertilizing Arabidopsis thaliana and its outcrossing relative Arabidopsis lyrata.** *BMC Evol Biol* 2010, **10:**10.
18. Quesneville H, Bergman CM, Andrieu O, Autard D, Nouaud D, Ashburner M, Anxolabehere D: **Combined evidence annotation of transposable elements in genome sequences.** *PLoS Comput Biol* 2005, **1:**166-175.
19. Flutre T, Duprat E, Feuillet C, Quesneville H: **Considering transposable element diversification in de novo annotation approaches.** *PLoS One* 2011, **6:**e16526.
20. Fransz PF, Armstrong S, de Jong JH, Parnell LD, van Drunen C, Dean C, Zabel P, Bisseling T, Jones GH: **Integrated cytogenetic map of chromosome arm 4S of A. thaliana: structural organization of heterochromatic knob and centromere region.** *Cell* 2000, **100:**367-376.
21. AGI AGI: **Analysis of the genome sequence of the flowering plant Arabidopsis thaliana.** *Nature* 2000, **408:**796-815.
22. Lippman Z, Gendrel AV, Black M, Vaughn MW, Dedhia N, McCombie WR, Lavine K, Mittal V, May B, Kasschau KD, et al: **Role of transposable elements in heterochromatin and epigenetic control.** *Nature* 2004, **430:**471-476.
23. WUGSC WUGSC: **The complete sequence of a heterochromatic island from a higher eukaryote. The Cold Spring Harbor Laboratory, Washington University Genome Sequencing Center, and PE Biosystems Arabidopsis Sequencing Consortium.** *Cell* 2000, **100:**377-386.
24. Tabata S, Kaneko T, Nakamura Y, Kotani H, Kato T, Asamizu E, Miyajima N, Sasamoto S, Kimura T, Hosouchi T, et al: **Sequence and analysis of chromosome 5 of the plant Arabidopsis thaliana.** *Nature* 2000, **408:**823-826.
25. Needleman SB, Wunsch CD: **A general method applicable to the search for similarities in the amino acid sequence of two proteins.** *J Mol Biol* 1970, **48:**443-453.
26. Guindon S, Gascuel O: **A simple, fast, and accurate algorithm to estimate large phylogenies by maximum likelihood.** *Syst Biol* 2003, **52:**696-704.
27. Petrov DA, Lozovskaya ER, Hartl DL: **High intrinsic rate of DNA loss in Drosophila.** *Nature* 1996, **384:**346-349.
28. Wicker T, Sabot F, Hua-Van A, Bennetzen JL, Capy P, Chalhoub B, Flavell A, Leroy P, Morgante M, Panaud O, et al: **A unified classification system for eukaryotic transposable elements.** *Nat Rev Genet* 2007, **8:**973-982.
29. Baucom RS, Estill JC, Chaparro C, Upshaw N, Jogi A, Deragon JM, Westerman RP, Sanmiguel PJ, Bennetzen JL: **Exceptional diversity, non-random distribution, and rapid evolution of retroelements in the B73 maize genome.** *PLoS Genet* 2009, **5:**e1000732.
30. Pereira V: **Insertion bias and purifying selection of retrotransposons in the Arabidopsis thaliana genome.** *Genome Biol* 2004, **5:**R79.

31. Sandmeyer SB, Hansen LJ, Chalker DL: **Integration specificity of retrotransposons and retroviruses.** *Annu Rev Genet* 1990, **24:**491-518.
32. Spradling AC, Stern DM, Kiss I, Roote J, Laverty T, Rubin GM: **Gene disruptions using P transposable elements: an integral component of the Drosophila genome project.** *Proc Natl Acad Sci U S A* 1995, **92:**10824-10830.
33. Gao X, Hou Y, Ebina H, Levin HL, Voytas DF: **Chromodomains direct integration of retrotransposons to heterochromatin.** *Genome Res* 2008, **18:**359-369.
34. Tenaillon MI, Hollister JD, Gaut BS: **A triptych of the evolution of plant transposable elements.** *Trends Plant Sci* 2010, **15:**471-478.
35. Petrov DA, Aminetzach YT, Davis JC, Bensasson D, Hirsh AE: **Size matters: non-LTR retrotransposable elements and ectopic recombination in Drosophila.** *Mol Biol Evol* 2003, **20:**880-892.
36. Virgin JB, Bailey JP: **The M26 hotspot of Schizosaccharomyces pombe stimulates meiotic ectopic recombination and chromosomal rearrangements.** *Genetics* 1998, **149:**1191-1204.
37. Montgomery E, Charlesworth B, Langley CH: **A test for the role of natural selection in the stabilization of transposable element copy number in a population of Drosophila melanogaster.** *Genet Res* 1987, **49:**31-41.
38. Bartolome C, Maside X, Charlesworth B: **On the abundance and distribution of transposable elements in the genome of Drosophila melanogaster.** *Mol Biol Evol* 2002, **19:**926-937.
39. Rizzon C, Marais G, Gouy M, Biemont C: **Recombination rate and the distribution of transposable elements in the Drosophila melanogaster genome.** *Genome Res* 2002, **12:**400-407.
40. Petrov DA, Fiston-Lavier AS, Lipatov M, Lenkov K, Gonzalez J: **Population genomics of transposable elements in Drosophila melanogaster.** *Mol Biol Evol* 2010.
41. Petrov DA, Fiston-Lavier AS, Lipatov M, Lenkov K, Gonzalez J: **Population Genomics of Transposable Elements in Drosophila melanogaster.** *Mol Biol Evol* 2011, **28:**1633-1644.
42. Duret L, Marais G, Biemont C: **Transposons but not retrotransposons are located preferentially in regions of high recombination rate in Caenorhabditis elegans.** *Genetics* 2000, **156:**1661-1669.
43. Nuzhdin SV: **Sure facts, speculations, and open questions about the evolution of transposable element copy number.** *Genetica* 1999, **107:**129-137.
44. Cam HP, Noma K, Ebina H, Levin HL, Grewal SI: **Host genome surveillance for retrotransposons by transposon-derived proteins.** *Nature* 2008, **451:**431-436.
45. Mosher RA, Melnyk CW: **siRNAs and DNA methylation: seedy epigenetics.** *Trends Plant Sci* 2010, **15:**204-210.
46. Hill WG, Robertson A: **The effect of linkage on limits to artificial selection.** *Genet Res* 1966, **8:**269-294.